\documentclass[prb,twocolumn,superscriptaddress, showpacs]{revtex4}

\usepackage[dvips]{graphicx}
\usepackage{amssymb,amsmath}

\pacs{74.72.-h, 75.10.JM }

\begin{document}
\title{Lightly doped dimerized spin chain in the
one-dimensional $t$-$J$-$J'$ model}
\author{Alexander Seidel}
\affiliation{Center for Material Science and Engineering, Massachusetts Institute of Technology, Cambridge, MA 02139}
\affiliation{Department of Physics, Massachusetts Institute of Technology, Cambridge, MA 02139}
\author{Patrick A. Lee} 
\affiliation{Center for Material Science and Engineering, Massachusetts Institute of Technology, Cambridge, MA 02139}
\affiliation{Department of Physics, Massachusetts Institute of Technology, Cambridge, MA 02139}
\date{\today}

\begin{abstract}
We study the one-dimensional $t$-$J$-$J'$-model in the limit of small hole doping $x$ and small $J/t$, $J'/t$. Special emphasis is put on the regime $J'/J \approx .5$ where a spin gap is present at small doping and the undoped spin chain is strongly dimerized. Using a perturbative approach and Luttinger liquid arguments, we demonstrate for this non-integrable class of models that the charge degrees of freedom behave as non-interacting spinless solitons in the dilute hole limit. Our approach is also used to evaluate the energy and mass renormalization of a single hole. Interestingly, the corrections of these quantities are in powers of $\sqrt{J/t}$. At $J'/J=.5$ we construct a variational spin-polaron wave function for the hole and find good agreement with our perturbative results. 
\end{abstract}
\maketitle

\section{Introduction}

\begin{figure}[hbt]
\samepage
\vspace{.2cm}
\includegraphics[width=3.3in]{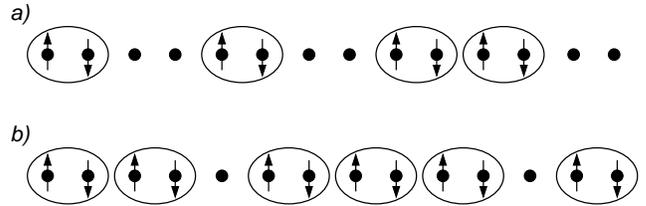}
\caption{\label{models} a) doping into frozen dimer state. Ovals represent singlet pairs. b) mobile dimers with domain walls.}
\vspace{-.5cm}
\end{figure}

\noindent The discovery of high-$T_c$ superconductors has generated great interest in models of strongly correlated insulators in low dimensions, both in the presence and absence of doped carriers. While in two dimensions the problem remains challenging even for the most elementary models that are believed to capture some of the physics of the cuprates, much can be learned from the study of one-dimensional counterparts of such models. Basic features that characterize the high-$T_c$ materials, such as the absence of a quasiparticle pole near the Fermi surface and possibly spin-charge separation, are essential to one-dimensional systems and are well understood here \cite{VOIT}. The success of the theory in one dimension is due to the availability of powerful exact methods, such as numerical diagonalization and Bethe ansatz, in combination with the knowledge of a low-energy effective theory which has so far been found to describe all gapless degrees of freedom in one-dimensional systems, the Luttinger liquid \cite{HAL1}.\\
\indent  In the present work we study the effect of doping a small concentration of holes into a dimerized spin chain. This is motivated by the idea that in a dimerized system, there naturally exists an amplitude for pair formation. Upon introduction of carriers phase coherence may be established, resulting in superconductivity. This idea has been previously proposed in the literature \cite{IMADA93}. We distinguish between two rough physical pictures of hole doping into a dimerized chain (Fig. \ref{models}). In the first scenario (Fig. \ref{models}a)) the dimer order remains long ranged even after hole doping. The dimers tend to reside on every second link of the lattice as in the symmetry broken undoped state, and holes tend to pair on the empty sites in between.
\begin{figure*}[t]
\samepage
\begin{center}
\includegraphics[width=6.5in]{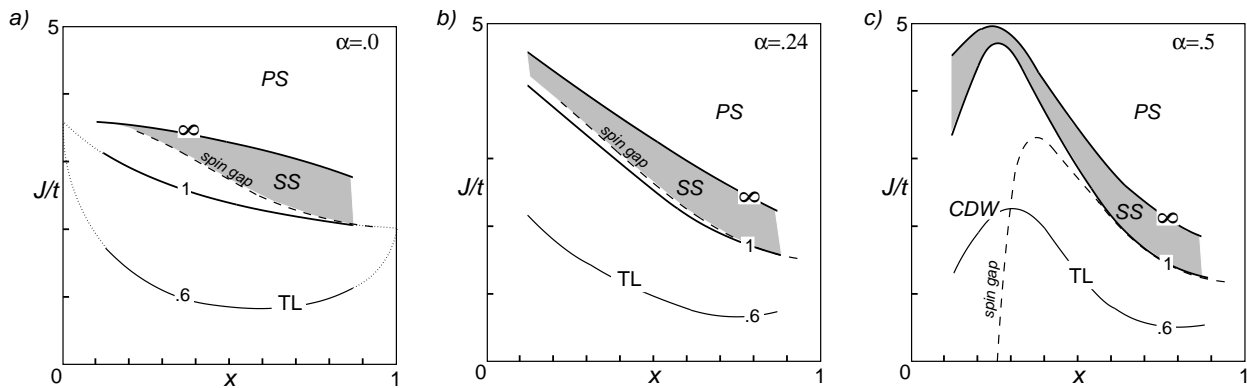}
\caption{\label{phas} Sketch of the zero temperature phase diagram of the $t$-$J$-$J'$-model as determined numerically in Ref.\onlinecite{NAKAMURA} for $\alpha=0$ (a), $\alpha=.24$ (b) and $\alpha=.5$ (c). Contours are labeled by values of $K_{\rho}$. The shaded region marks the domain of dominant singlet superconducting correlations. The dotted lines in (a) were proposed in Ref. \onlinecite{OGSOAS}.}
\vspace{-.4cm}
\end{center}
\end{figure*}
We note that this case bears some resemblance to doped ladder models \cite{DARI}. Indeed strong superconducting correlations have been proposed for dimer-models with explicit symmetry breaking \cite{IMADA, TANXU}. \\
\indent A second possibility is that the holes enter as domain walls between different dimer phases (Fig.\ref{models}b). Then the long-range order of dimers is destroyed, but the singlet gap remains. If the holes are mobile and become phase coherent, this leads to superconductivity in analogy with the original RVB scenario \cite{PWA}. However, the pictures presented in Fig. \ref{models} are based on the limit of large $J/t$. In physical systems, on the other hand, $J/t$ is usually a fraction of unity. In this regime, the hole kinetic energy is dominant and it is less clear intuitively if the dimer picture discussed above applies. In the present paper, we study this question in the context of the $t$-$J$-$J'$-model in one dimension:
\begin{equation}\label{tJJ'}
\begin{split}
H& =-t P\sum_i(c^\dagger_{i,\sigma}c_{i+1,\sigma}+h.c.)P 
+J\sum_i(S_i\cdot S_{i+1}-\frac{1}{4}n_in_{i+1}) \\
&\quad +J'\sum_i(S_i\cdot S_{i+2}-\frac{1}{4}n_in_{i+2}) +V\sum_i{n_in_{i+1}}
\end{split}
\end{equation}
Here, $P$ is a projection operator that enforces the constraint of no doubly occupied sites. The $S_i$ are spin-$1/2$ operators, and the nearest- and next-nearest-neighbor couplings $J$ and $J'$ are assumed positive throughout this paper. In addition, we have included a nearest neighbor interaction $V$ for later convenience.

At finite doping, not much is known analytically about this model due to its non-integrable nature. The phase diagram has been established numerically \cite{OGSOAS, OGLURI, NAKAMURA} (Fig. \ref{phas}). However, in the regime around $\alpha\equiv J'/J\approx .5$ where the undoped spin chain is strongly dimerized and which is of particular interest to us, we feel that the numerics are somewhat inconclusive for small doping $x$ and $J/t\lesssim 1$ as we will discuss below.

\indent The purpose of this paper is twofold: First,  in the case of strong frustration $\alpha\equiv J'/J\approx .5$ we wish to determine the fate of the various regions present in the phase diagram of the $t$-$J$-$J'$-model at small doping and small $J/t$ (Fig.\ref{phas}c). In particular, we will answer the question whether a regime of dominant singlet superconducting correlations in the vicinity of an instability towards phase separation persists to values $J/t\ll 1$ and in the dilute hole limit. This would happen if all the contours in Fig. \ref{phas}c) extrapolate to the origin, which appears to be a possible interpretation of the numerics. To this end, we introduce a perturbative approach valid in this limit. We find that the scenario mentioned above does not occur, but instead the Luttinger liquid is stable for small $J/t$ and small doping. Luttinger liquid arguments will then imply that strong superconducting correlations only exist above a finite critical value of $J/t$. \\
\indent Second, below the critical value $(J/t)_c$ the liquid phase is stable in the limit $x\rightarrow 0$, and we will use our method to demonstrate certain properties that one expects to hold for one-dimensional lattice models based on general grounds. In particular, spin and charge are expected to correspond to separate degrees of freedom and any microscopic coupling between them should be irrelevant. As a consequence, dilute holes that are doped into a correlated spin chain should act as a gas of non-interacting spinless solitons\cite{HALCOM}, where the coupling to the non-trivial spin background only gives rise to a renormalization of the effective hole mass. This phenomenon has been observed in integrable models such as the Hubbard model \cite{SCHULZ}. Our perturbative approach allows us to give a demonstration of the same behavior in the non-integrable $t$-$J$-$J'$-model, and to calculate the effective energy and mass renormalization of the hole for small $J/t$.\\
\indent The remainder of this paper is organized as follows: In section \ref{model} we introduce the model, briefly discuss the numerical phase diagram and cast the model into a language where holes are interpreted as domain walls. In section \ref{pert} we treat the spin-charge couplings as a perturbation and derive expressions for the ground-state energy, compressibility and Kohn stiffness of the model for small $J/t$, $J'/t$ and small doping $x$. This will allow us to qualitatively continue the numerical phase diagram into the region of small doping. In section \ref{eval} we will explicitly evaluate these expressions as asymptotic series in powers of $x$ and $\sqrt{J/t}$, demonstrating the convergence of our approach. In section \ref{polaron} we discuss the single polaron energy and mass renormalization. We introduce a spin polaron type variational wave function where the hole is surrounded by a cloud of tightly bound triplet excitations and find good agreement with our perturbative results. We conclude in section \ref{conclusion}. Appendix \ref{convapp} illustrates the behavior of our expansion at general order in various limits. An important technical issue is discussed in appendix \ref{Fcontin}.

\section{\label{model} Formulation of the problem}
\begin{figure}[hbt]
\samepage
\includegraphics[width=3.3in]{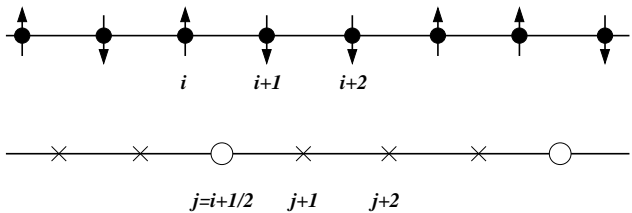}
\caption{\label{cluster} Domain wall representation of the Hilbert space. The upper chain represents the spin sector with sites labeled by i. The holes live on a lattice of interstitial sites labeled by j. Crosses represent ``empty'' interstitial sites, circles represent holes between two spins.}
\end{figure}

We wish to study the Hamiltonian (\ref{tJJ'}) in the limit of vanishing exchange couplings and doping.
For $V=0$ the phase zero temperature diagram of (\ref{tJJ'}) has been obtained numerically \cite{OGSOAS, OGLURI, NAKAMURA} for various values of the parameter $\alpha\equiv J'/J$. The results are sketched in Fig. \ref{phas}). These phase diagrams show a Tomonaga-Luttinger liquid (TL) region below the dashed line labeled spin gap, where both spin and charge degrees of freedom are gapless. Above the dashed line there is a spin gapped liquid phase which is subdivided into a regime of dominant singlet superconducting (SS) correlations (shaded grey) and, where present, a regime of dominant charge-density-wave correlations (CDW). Also shown are contours of constant values of the Luttinger parameter $K_\rho$. Above the $K_\rho$ line labeled $\infty$, there is a region of phase separation (PS), where the ground state has a phase boundary between a hole-rich and an electron-rich phase. The parameter $K_\rho$ is directly related to the large distance behavior of the various correlation functions of a Luttinger liquid:  For $K_\rho>1$ pairing correlations dominate over density-wave correlations, otherwise density-wave correlations dominate. Furthermore, in the presence of a spin gap, triplet pairing correlations and spin-density-wave correlations are exponentially suppressed. In this region, the main competition is therefore between CDW and SS correlations, which in the presence of a spin gap decay as $r^{-K_\rho}$ and $r^{-1/K_\rho}$ respectively \cite{SOLYOM, VOIT}.\\
\indent At zero doping ($x=0$) it is well known that the spin chain undergoes a phase transition at a critical value of $\alpha_c\approx .24$ (Ref. \onlinecite{JULHAL}) above which the ground state is dimerized with a gap in the spin excitation spectrum \cite{HAL3}. For $\alpha>\alpha_c$ this spin gap remains present over a finite range of doping for any value of $J/t$, as was shown in Ref.\onlinecite{OGLURI}. Fig. \ref{phas} shows that at $\alpha=.5$ the spin gapped region has considerable overlap with the pairing region $K_\rho>1$ even at small values of doping $x$. Furthermore, it appears from the numerics as if all $K_\rho$ contours, including the phase separation boundary $K_\rho=\infty$, flow to small values of $J/t$ at small $x$ (cf. Refs. \onlinecite{OGLURI, NAKAMURA}). One expects that these contours will focus on a critical point $[x=0,(J/t)_c]$, as was proposed in Ref. \onlinecite{OGSOAS} for the $t$-$J$-model ($J'=0$, Fig. \ref{phas}a)). A possibility that seems consistent with the numerics at $\alpha=.5$ is that $(J/t)_c=0$, i.e. all contours flow into the origin of the phase diagram. In this case, a sufficiently small amount of doping would always lead to phase separation, and upon further doping one would enter a region of dominant SS correlations. Alternatively, $(J/t)_c$ could be finite but possibly smaller than its value at $\alpha=0$, which is between $3$ and $4$ (Fig. \ref{phas}a). This would imply that the above phenomenology of phase separation and superconductivity at small doping occurs only for $J/t>(J/t)_c$, while for $J/t<(J/t)_c$ the liquid phase is stable at any doping $x$. In the latter case, one would expect the Luttinger parameter $K_\rho$ to approach the value $1/2$ in the dilute hole limit $x\rightarrow 0$, which is the value corresponding to non-interacting spinless degrees of freedom. This behavior is clearly exemplified by the numerical phase diagram of the $t$-$J$-model (Fig. \ref{phas}a). The primary goal of this paper is to determine which of these two scenarios applies to the $t$-$J$-$J'$ model at $\alpha>\alpha_c$.\\
\indent We begin our analysis by casting the Hamiltonian (\ref{tJJ'}) into a language where the holes play the role of domain walls between broken segments of an infinite spin chain \footnote{We thank F.D.M. Haldane for pointing out to us this possibility} (Fig. \ref{cluster}). We consider a lattice of $L$ sites with a number of $N_e$ electrons and $N_h=L-N_e$ holes. 
Denoting the $i$'th spin on the lattice by $S_i$ we may regard the spins as residing on a ``squeezed'' lattice where the hole sites have been dropped from the system and the label $i$ of the spin $S_i$ is a site label in this squeezed space, as in Fig. \ref{cluster}). We also introduce interstitial sites for the squeezed spin lattice whose labels $j=i+1/2$ differ from those of the spin sites by $1/2$. Each interstitial site may accommodate a number $n_j=0,1...$ of holes. A faithful representation of the Hilbert space of (\ref{tJJ'}) is given by states labeled by
\begin{equation}\label{states}
\begin{split}
 \left|\dotsc,\sigma_i,n_{i+\frac{1}{2}},\sigma_{i+1},\dotsc\right>&\equiv\left|\dotsc\sigma_i\dotsc\right>\left|\dotsc n_{j}\dotsc\right>\\
   \sum_j{n_j}&=N_h\\
    i\, &=1\dotsc N_e \\ 
    j\, &=\frac{1}{2}\dotsc N_e+\frac{1}{2}
\end{split}
\end{equation} 
\noindent where $\sigma_i=\pm\frac{1}{2}$ denotes the z-component of the spin $S_i$. This language turns out to be particularly convenient when one introduces a large nearest neighbor hole repulsion, i. e. if we let $V=\infty$ in (\ref{tJJ'}), such that it is forbidden for two holes to occupy neighboring sites. In the present language this means that the occupancy of the hole cluster labeled by $j$ is now restricted to be $n_j=0,1$. This modification of the model will be irrelevant in the dilute hole limit which we are interested in\footnote{Note that for small doping and $J/t$ the (PS) region consists of an insulating $x=0$ phase (no holes) and a liquid phase with $x=x_c$, where $x_c$ is the critical doping at which phase separation just occurs. Since $x_c$ is small in the region of interest, it is justified to consider the holes as dilute even in the (PS) regime.}.
We may now choose to formulate the hole dynamics either in terms of hard-core boson operators or spinless fermion operators. For convenience, we introduce fermion ladder operators $c_j$, $c^\dagger_j$, where the action of $c^\dagger_j$ can be described by cutting the spin chain open at the interstitial site $j$, introducing a hole at this position and multiplying the state by an appropriate fermion phase. The hole kinetic energy is then simply given by

\begin{equation}\label{Hc}
H_c=-\sum_j\left(t\,c^\dagger_jc_{j+1}+h.c.\right)
\end{equation} 

\noindent (\ref{Hc}) can be thought of as the $J=J'=0$ limit of the Hamiltonian. In the other limit of interest, namely the limit of zero doping $x=N_h/L$, the Hamiltonian becomes that of a pure spin chain:

\begin{equation}\label{Hs}
\begin{split}
H_s&=J\sum_i\left(X_{i,i+1}+\alpha X_{i,i+2}\right)\\
\text{where}\quad\quad X_{i,i'}&=S_i\cdot S_{i'}-\frac{1}{4}\\
\end{split}
\end{equation} 
 
\noindent where we work at constant $\alpha=J'/J$ from now on, and assume that $\alpha>\alpha_c$, such that the small doping regime is spin gapped. The combined Hamiltonian 

\begin{equation}\label{H0}
H_0=H_s+H_c,
\end{equation}  

\noindent where the spin and charge part are still completely decoupled, will serve as a starting point for the perturbation theory we propose. In order to correctly reproduce matrix elements of the Hamiltonian (\ref{tJJ'}), couplings between the spin sector and the charge sector must be introduced:

\begin{align}
H&=H_0+H_{sc}+H'_{sc} \label{H}\\
H_{sc}&=-J\sum_jn_j\gamma_j \label{Hsc}\\
H'_{sc}&=J\alpha\sum_i X_{i-1,i+1}n_{i-\frac{1}{2}}n_{i+\frac{1}{2}}\label{Hsc'}\\
\text{where}&\nonumber\\
\gamma_{i+\frac{1}{2}}&=(1-\alpha)X_{i,i+1}+\alpha(X_{i-1,i+1}+X_{i,i+2})
\end{align}  

\noindent Here, $H_{sc}$ is a correction which couples spin and charge by adjusting nearest neighbor bonds and removing next-nearest neighbor bonds in the squeezed spin space in the vicinity of a hole. Certain corrections of the latter sort are redundant whenever two holes are next-nearest neighbors in real space -- or nearest neighbors in the present formalism -- and this is corrected by $H'_{sc}$. Formally, $H_{sc}$ and $H'_{sc}$ are suppressed by powers of both $J/t$ {\em and} $x$, and hence can be regarded as small compared to $H_0$. Our strategy is thus to treat the spin-charge coupling terms $H_{sc}+H_{sc}'$ as a perturbation. We must caution, however, that the hole kinetic energy $H_c$ is very small, or order $tx^2$, and the small energy denominators that appear in perturbation theory must be treated with care. We apply this method to the spin gapped regime $\alpha>\alpha_c$ and find it to be a valid procedure in second order perturbation theory, in the sense that corrections are indeed small in the limit we consider. The general systematics of this at higher order perturbation theory are elucidated in appendix \ref{convapp}.\\
\indent In the following section we will show that our approach gives rise to a perturbative expansion of the ground-state energy which may be used to analyze the phase diagram of (\ref{tJJ'}) in the vicinity of the origin, where the fate of the various phases is uncertain from numerics for $\alpha>\alpha_c$ (Fig.\ref{phas}c). We note that the procedure proposed here bears some resemblance to that used by Xiang et al. to study the $t$-$J$-model in a first order perturbative approach \cite{XIANG}.

\section{\label{pert}Perturbative analysis of the model}

\noindent In gapless one-dimensional systems it is generally possible to derive basic features of the phase diagram from spectral properties by means of Luttinger liquid theory \cite{HAL1}. The low-energy properties of 
a Luttinger liquid are completely defined in terms of three parameters which have the dimension of a velocity: A sound velocity $v_s$, a ``compressibility'' parameter $v_N$ related to elementary charge excitations, and a ``stiffness'' parameter $v_J$ related to elementary current excitations. These are not independent, but are related by the following universal relations identified by Haldane \cite{HAL1}:
\begin{equation}
 v_N=v_s/K_\rho , \hspace{.5cm} v_J=v_s K_\rho 
\label{velos}
\end{equation}

\noindent The parameters $v_N$ and $v_J$ can be calculated from the dependence of the ground-state energy $E_0$ on the carrier density $x$ and on a phase twist $\phi$ respectively \cite{VOIT}, where $\phi$ is associated with a flux $L\phi$ penetrating the system when it is imposed on a ring with periodic boundary conditions:

\begin{align}
 v_N &= \frac{2}{\pi L}\frac{\partial^2 E_0}{\partial x^2}\label{vn}\\
 v_J &= \left.\frac{\pi}{2L}\frac{\partial^2 E_0}{\partial\phi^2}\right|_{\phi=0}\label{vj}
\end{align}

By (\ref{vn}), (\ref{vj}), $v_N$ is proportional to the inverse compressibility of the system, while $v_J$ is proportional to the Kohn stiffness\cite{Kohn}, which is related to the Drude weight of the conductivity. Eqs. (\ref{velos})-(\ref{vj}) allow the determination of $K_\rho$ via

\begin{equation}
 K_\rho=\sqrt{\frac{v_J}{v_N}} 
\label{Krho}
\end{equation}
 
\noindent from the ground-state energy of the system alone. The strategy in now to evaluate both the numerator and the denominator in (\ref{Krho}) perturbatively.\\
\indent We now proceed by imposing periodic boundary conditions on the charge sector and the spin sector of the system (\ref{H})-(\ref{Hsc'}) separately. This is apparently not the same as imposing periodic boundary conditions in real space, since momenta are now quantized in units of $2\pi/N_e$ rather than $2\pi/L$. Note that there is a unique and well defined map between the state space introduced in (\ref{states}) and the Hilbert space of the $t$-$J$-$J'$ model only for a finite system with {\em open} boundary conditions. Indeed, imposing open boundary in real space {\em is} equivalent to imposing them in the spin sector and the charge sector separately. However, going from open to periodic boundary conditions is not expected to matter for large system sizes. The unperturbed Hamiltonian $H_0$ then has two separately conserved momenta, and we denote its ground state by 

\begin{equation}\label{gs}
\left|\sigma_0,\psi_0\right>\equiv\left|\sigma_0\right>\left|\psi_0\right>
\end{equation}

\noindent where $\left|\sigma_0\right>$ is the ground state of the spin Hamiltonian $H_s$ on a ring of $N_e$ spin sites. Although for $\alpha>\alpha_c$ the ground state of $H_s$ has a broken translational symmetry and is doubly degenerate, we will assume that $\left|\sigma_0\right>$ is a symmetric superposition of the two symmetry broken ground states and thus has zero lattice momentum. Likewise $\left|\psi_0\right>$ is a non-interacting Fermi sea of $N_h$ spinless Fermions hopping on $N_e$ sites with periodic boundary conditions. The unperturbed ground-state energy we write as

\begin{align}
 E_0&=E_{\sigma_0}+E_{\psi_0}=E_{\sigma_0}-N_e\frac{2t}{\pi}\sin(k_f)\label{E00}\\
\text{where} \quad k_f&=\pi\frac{N_h}{N_e}=\frac{\pi x}{1-x}\label{kf}  
\end{align}

\noindent and $E_{\sigma_0}$, $E_{\psi_0}$ are the ground-state energy of $H_s$ and the spinless fermion kinetic energy, respectively. We will focus our analysis on the limit $x\rightarrow 0$ where $J/t$ is small but fixed. In this limit we argue that the ground-state energy of (\ref{H}) has an asymptotic expansion of the form 

\begin{equation}
\label{Ex}
E=E_{\sigma_0}+L(Ax +Bx^2 +Cx^3+\dotsc)
\end{equation}

\noindent The coefficients $A,B\dotsc$ will depend on $J/t$. At the leading order, they can in principle be inferred from the spinless fermion kinetic energy in (\ref{E00}). Formally, however it will be more convenient to work with an expansion of the form 

\begin{equation}
\label{Ekf}
E=E_{\sigma_0}+N_e\left(\tilde A\left(\frac{k_f}{\pi}\right)+\tilde B\left(\frac{k_f}{\pi}\right)^2+\tilde C\left(\frac{k_f}{\pi}\right)^3 +\dotsc\right)
\end{equation}

\noindent The coefficients in (\ref{Ex}) and (\ref{Ekf}) will in general not be the same, due to the non-linear dependence of $k_f$ on $x$ in (\ref{kf}). However, since $N_e=L(1-x)$, the $\tilde A$ term in (\ref{Ekf}) is linear in $x$, and hence

\begin{equation}
\label{AB}
A=\tilde A, \quad B=\tilde B
\end{equation}

\noindent We will now proceed by evaluating the above series order by order in perturbation theory, treating the spin-charge coupling terms $H_{sc}+H'_{sc}$ perturbatively as we have outlined above. We write

\begin{equation}\label{orders}
\begin{split}
E&=E_0+E_1+E_2+\dotsc+E_k+\dotsc \\
A&=A_0+A_1+A_2+\dotsc+A_k+\dotsc\\
&\;\;\vdots
\end{split}
\end{equation}

\noindent and similarly for all other coefficients, where the label $k$ denotes a term arising at $k$'th order perturbation theory. We have:

\begin{equation}\label{zero}
A_0=-2t \quad B_0=0 \quad C_0= \frac{\pi^2}{3}t
\end{equation} 

\noindent From (\ref{vn})-(\ref{Krho}), (\ref{Ex}) it follows that at small $J/t$

\begin{equation}
\label{Kpert}
K_\rho=\frac{\pi}{2} \sqrt{\frac{A_{\phi\phi}x}{2B+6Cx}}\approx \frac{1}{2}\sqrt{\frac{tx}{B/\pi^2+tx}}
\end{equation}

\noindent where $A_{\phi\phi}$ denotes the second derivative with respect to the phase $\phi$ introduced above, and $A_0(\phi)=-2t\cos(\phi)$ was used. Eq. (\ref{Kpert}) shows that if $B$ acquires a finite negative value due to the spin-charge couplings, $K_\rho$ will diverge as $x\rightarrow 0$ even at small $J/t$. This would imply strong superconducting fluctuations and phase separation at the divergence. On the other hand, if $B$ is zero or positive, the liquid phase will be stable for small $x$ and $J/t$, and dominant superconducting correlations will be absent in the vicinity of the origin of the phase diagram. In this case, one would expect that $B=0$ to all orders in perturbation theory, since for $B>0$ one would have $K_\rho\rightarrow 0$ as $x\rightarrow 0$ which seems inconsistent with the numerical phase diagram. Also, $K_\rho=0$ is a somewhat unlikely pathological limit of Luttinger liquid theory, where the coefficient of the conjugate momentum of the charge field vanishes and the charges freeze into a classical state. This again seems unlikely in the absence of long-rage interactions, which one may assume especially in the presence of a spin gap. As we have argued initially, one would rather expect the charges to behave as non-interacting spinless solitons in the dilute limit, and Luttinger liquid physics then implies that $K_\rho$ assumes the value $\frac{1}{2}$ in this limit, provided that no instability intervenes.
This then requires that the coefficient $A_{\phi\phi}$ and $C$ in (\ref{Kpert}) are not independent, but have a constant ratio independent of $J$.\\
\indent Hence, we distinguish only two cases which we feel are the only ones consistent with the numerical phase diagram for $\alpha>\alpha_c$, Luttinger liquid theory and general expectations from the study of integrable models\cite{HAL1,SCHULZ}: 1. $B<0$, leading to a phase separation instability in the dilute hole limit for any value of $J/t$, and 2. $B\equiv 0$, $6C\equiv\pi^2A_{\phi\phi}$ for any $J$, corresponding to a stable liquid as $x\rightarrow 0$ and small $J/t$, where the charges act as a dilute gas of non-interacting spinless solitons. We will now show that the second case applies.\\
\indent At first order perturbation theory the energy corrections factorize into mean-field like products, since spin and charge are not correlated in the ground state wave function (\ref{gs}). We have

\begin{align}
E_1&= \left<\sigma_0, \psi_0\left|H_{sc}+H_{sc}'\right|\sigma_0, \psi_0\right>\nonumber\\
&=-N_eJ\left<n_j\right>_0\left<\gamma_j\right>_0+N_eJ\alpha\left<X_{i-1,i+1}\right>_0\left<n_jn_{j+1}\right>_0\nonumber\\
&=-L\left<\gamma_j\right>_0\,Jx + LJ\,O(x^4)\\
\text{hence:}\quad\quad A_1&=-\left<\gamma_j\right>_0J \label{A1}
\end{align}

\noindent where $<>_0$ denotes the expectation value with respect to $|\sigma_0>$ or $|\psi_0>$ when no ambiguity is possible. Note that the contribution of $H'_{sc}$ is of order $x^4$. The smallness of this term as $x\rightarrow 0$ reflects the fact that the holes obey the Pauli principle which suppresses the probability of two holes being near each other. We see that already at this order, $H'_{sc}$ does not renormalize any of the coefficients in (\ref{Ex}) that we are interested in. We will thus drop it from the subsequent discussion. To determine leading corrections to $B$ and $C$, we will need to go to second order:

\begin{equation}\label{E2}
 E_2= -\sum_{\left|\sigma,\psi\right>}'\frac{\left<\sigma_0,\psi_0\right|H_{sc}\left|\sigma,\psi\right>\left<\sigma,\psi\right|H_{sc}\left|\sigma_0,\psi_0\right>}{E_\psi-E_{\psi_0}+E_\sigma-E_{\sigma_0}}
\end{equation}

\noindent where the sum goes over a complete set of unperturbed eigenstates and the prime excludes the ground state (\ref{gs}) from the sum. We now rewrite $H_{sc}$ as

\begin{equation}\label{HscFT}
H_{sc}=-\frac{J}{N_e}\sum_q n_q\gamma_{-q}
\end{equation}

\noindent where Fourier transforms

\begin{equation}
\begin{split}
 n_q&\equiv \sum_j e^{iqj}n_j=\sum_k c^\dagger_{k+q}c_k\\
\gamma_q&\equiv \sum_j e^{iqj}\gamma_j
\end{split}
\end{equation}

\noindent have been introduced. Using the fact that the intermediate states in (\ref{E2}) can be chosen to be momentum eigenstates, we have

\begin{equation}\label{E2_2}
 E_2= -\frac{J^2}{N_e^2}\sum_q\sum_{\left|\sigma,\psi\right>}'
\frac{
\left<\psi_0\right|n_{-q}\left|\psi\right>\left<\psi\right|n_q\left|\psi_0\right>
\left<\sigma_0\right|\gamma_q\left|\sigma\right>\left<\sigma\right|\gamma_{-q}\left|\sigma_0\right>
}{E_\psi-E_{\psi_0}+E_\sigma-E_{\sigma_0}}
\end{equation}

\noindent It is necessary to distinguish between terms with zero momentum exchange between spin and charge and those with $q\neq 0$. We write:

\begin{equation}\label{Esplit}
 E_2=E_2^0+E'_2
\end{equation}

\noindent where $E_2^0$ contains all $q=0$ terms and $E'_2$ contains all the rest. At $q=0$, $n_q=\sum{n_j}=N_h$ commutes with the Hamiltonian, hence there can be no virtual charge excitation and the charge matrix element is diagonal, $|\psi>=|\psi_0>$:

\begin{equation}\label{E0}
  E_2^0= -\left(\frac{k_f}{\pi}\right)^2J^2\sum_\sigma'{\frac{\left<\sigma_0\right|\gamma_{q=0}\left|\sigma\right>\left<\sigma\right|\gamma_{q=0}\left|\sigma_0\right>}{E_\sigma-E_{\sigma_0}}}
\end{equation}

\noindent Note that virtual states without spin excitations do not enter (\ref{E2_2}), since  $|\sigma>=|\sigma_0>$ would imply $q=0$ and again the charge part vanishes unless also $|\psi>=|\psi_0>$, which is excluded from the sum. Thus for $\alpha>\alpha_c$ the energy denominator in (\ref{E2_2}) is bounded from below by the spin gap $\Delta$ which will dominate over charge excitation energies of order $tx^2$ very close to the Fermi surface. This assures that the perturbative expansion is well behaved in the limit $x\rightarrow 0$ (see appendix \ref{convapp}). \\
\indent For $q\neq 0$ we note that $n_q$ excites only single particle-hole excitations. We can thus convert the sum over these terms into a double integral over a hole momentum $k_1$ and a particle momentum $k_2$:

\begin{align}\label{E'}
E'_2&=-N_e\int_{-k_f}^{k_f}\frac{dk_1}{2\pi}\int_{k_f}^{2\pi-k_f}\frac{dk_2}{2\pi}\;f(k_1,k_2)\\
\text{where}\quad\quad&  \nonumber\\
f(k_1,k_2)&=\frac{J^2}{N_e}\sum_{\left|\sigma\right>}'\frac{\left<\sigma_0\right|\gamma_{k_2-k_1}\left|\sigma\right>\left<\sigma\right|\gamma_{k_1-k_2}\left|\sigma_0\right>}{\epsilon(k_2)-\epsilon(k_1)+E_\sigma-E_{\sigma_0}} \label{f}
\end{align}

\noindent and $\epsilon(k)=-2t\cos(k)$ is the free fermion dispersion.
For later convenience, we also introduce the function

\begin{equation}\label{F}
\begin{split}
  F(q)&=\frac{1}{2}\left(f(\frac{q}{2},-\frac{q}{2})+f(-\frac{q}{2},\frac{q}{2})\right)\\
   &=\frac{J^2}{2N_e}\left(\sum_{\left|\sigma\right>}'\frac{\left<\sigma_0\right|\gamma_{-q}\left|\sigma\right>\left<\sigma\right|\gamma_{q}\left|\sigma_0\right>}{E_\sigma-E_{\sigma_0}}+(q\rightarrow -q)\right)
\end{split}
\end{equation}

\noindent where the symmetry of $\epsilon(k)$ was used.

The leading correction to the energy at second order perturbation theory is a contribution to the $A$ coefficient in (\ref{Ex}):

\begin{equation}\label{A2}
\begin{split}
  A_2&=\left.\frac{\pi}{N_e}\frac{\partial}{\partial k_f}E'_2\,\right|_{k_f=0}\\
     &=-\int_0^{2\pi}\frac{dk}{2\pi}\;f(0,k)
\end{split}
\end{equation}

\noindent To leading order in $J/t$ the integral over momenta may be carried out to give a quantity defined in terms of the pure spin chain $H_s$. This will demonstrate that the present expansion is well behaved, but we defer the evaluation to the following section in order to continue with the analysis of the crucial $B$ coefficient. Its correction at this order reads:
\begin{equation}\label{B2}
\begin{split}
  B_2&=\left.\frac{\pi^2}{2N_e}\left(\frac{\partial^2}{\partial k_f^2}E^0_2+\frac{\partial^2}{\partial k_f^2}E'_2\right)\,\right|_{k_f=0}\\
     &=\left.-F(0)+\frac{\pi^2}{2N_e}\frac{\partial^2}{\partial k_f^2}E'_2\,\right|_{k_f=0}
\end{split}
\end{equation}
\noindent Again, the contribution from $E'_2$ is evaluated by straightforward differentiation of (\ref{E'}). Only boundary terms survive, as all derivatives of the integrand vanish by symmetry when the limit $k_f\rightarrow 0$ is taken, using the $2\pi$-periodicity of $f$ in the second argument. We find:
\begin{equation}\label{B2_2}
\begin{split}
  B_2&=-F(0)+\left.\frac{1}{2}\left( F(2k_f)+F(2\eta)\,\right)\right|_{k_f\rightarrow 0}
\end{split}
\end{equation}
\noindent At this point we have introduced an infinitesimal $\eta$ since terms with zero momentum transfer are really excluded in the sum defining $E'_2$. However, we argue that the function $F(q)$ will be continuous at $q=0$ and hence $B_2$ vanishes. We note that this is the effect of a non-trivial cancellation between $q=0$ processes and processes with $q\rightarrow 0$. Physically, the continuity of $F(q)$ can be seen by interpreting $F(q)$ as the second order energy response of a pure spin chain due to a periodic perturbation, as we explain in detail in appendix \ref{Fcontin}. Hence, by (\ref{B2_2})

\begin{equation}\label{B2=0}
  B_2=0
\end{equation}

\noindent We have convinced ourselves that a similar cancellation in the $B$ coefficient takes place at third order perturbation theory\cite{THESIS}. We therefore propose that

\begin{equation}\label{B=0}
  B=0
\end{equation}

\noindent to all orders in perturbation theory, and thus for small $J/t$ the liquid remains stable in the limit $x \rightarrow 0$ even in the case $\alpha\approx .5$ (Fig. \ref{phas}c). The physical implication of (\ref{B=0}) is that indeed the holes act as spinless fermions whose interaction is short ranged, and is irrelevant in the dilute limit. The Pauli principle severely suppresses the wave function when two holes approach each other. The range of this suppression is larger in one dimension than for dimensions greater than one, since in higher dimensions a curvature of the wave function is less costly at small distances. Therefore, in one dimension this effect is strong enough in order to prevent a short range interaction from generating a term of order $x^2$ in the energy. \\
\indent As we have argued above, the non-interacting nature of the charge degrees of freedom in combination with Luttinger liquid arguments also imposes constraints on the linear and cubic terms in $x$ when a magnetic flux is imposed.
We now move on to verify these relations perturbatively. Note that $B=0$ leads to $C=\tilde C$ in (\ref{Ex}) and (\ref{Ekf}), such that

\begin{equation}\label{C2}
 C_2 = \left.\frac{\pi^3}{6N_e}\frac{\partial^3}{\partial k_f^3}E'_2 \right|_{k_f=0}  
\end{equation}

\noindent In this case boundary terms such as those displayed in (\ref{B2_2}) do not contribute, since they vanish by symmetry as $k_f \rightarrow 0$ when another derivative is acting on them. Instead, we have now again a ``bulk'' contribution analogous to that in (\ref{A2}):

\begin{equation}\label{C2_2}
 C_2 = -\left.\frac{1}{6}\,\frac{\pi}{2}\int_0^{2\pi}dk\;\frac{\partial^2}{\partial k_f^2}f(k_f,k) \right|_{k_f=0}  
\end{equation}

\noindent As explained above, this is to be compared to the coefficient $A_{2,\phi\phi}$. The phase twist $\phi$ will modify all hopping matrix elements via $t\rightarrow t\,e^{i\phi}$ in (\ref{Hc}) and leads to the following replacement of the free hole dispersion in the function $f(k_1,k_2)$ in (\ref{f}):

\begin{equation}
  \begin{split}
   \epsilon(k)&\longrightarrow \quad\epsilon_\phi(k)=-2t\cos(k+\phi)\\
   f(k_1,k_2)&\longrightarrow f_\phi(k_1,k_2)
  \end{split}
\end{equation}

\noindent Hence from (\ref{A2})

\begin{equation} \label{A2phi}
   A_{2,\phi,\phi}=\left.\frac{\partial^2}{\partial\phi^2}A_2\right|_{\phi=0}
     = -\int_0^{2\pi}\frac{dk}{2\pi}\;\left.\frac{\partial^2}{\partial\phi^2}f_\phi(0,k)\right|_{\phi=0}
\end{equation}

\noindent However, using the fact that

\begin{equation}
   f_\phi(0,k)=f(\phi,k+\phi)
\end{equation}
 
\noindent holds, it follows by shifting the integration variable and comparison with (\ref{C2_2}) that 

\begin{equation}
   6C_2 = \pi^2 A_{2,\phi\phi}
\end{equation}
 
\noindent is satisfied. Again, we have confirmed an identical relation at third order perturbation theory\cite{THESIS}, and this suggests that indeed
\begin{equation} \label{Aphi}
   6C = \pi^2 A_{\phi\phi}
\end{equation}
\noindent to all orders. Hence, although the parameters $v_N$ and $v_J$  in (\ref{Krho}) each receive nontrivial corrections, their ratio is fixed to leading order in $x$ such that $K_\rho$ always approaches $\frac{1}{2}$ in the limit $x\rightarrow 0$. Luttinger liquid theory then implies that the dilute holes share all the universal properties of a gas of non-interacting spinless particles. Presumably, this picture will hold in the entire regime $J/t<(J/t)_c$, where for $J/t>(J/t)_c$ small doping will give rise to phase separation.\\

\section{\label{eval} Explicit evaluation of coefficients}

\noindent In the preceding section we have shown that our perturbative approach is consistent in all details with a picture where the charge degrees of freedom behave as non-interacting spinless solitons in the dilute limit, and are effectively decoupled from the spin dynamics. The second order expressions we derived involve complicated sums over both spin and charge degrees of freedom. We will now show that the expressions for $A_2$ and $C_2$ can be evaluated more explicitly, to leading order in $J/t$, in terms of quantities that are derived from a pure spin chain problem. In this way we obtain explicit asymptotic expansions for the ground-state energy and the compressibility parameter $v_N$, showing that second order corrections are suppressed by non-trivial powers of $J/t$ compared to the leading orders. Also, these quantities are related to the single hole energy and mass renormalization, which will be clarified in the following section.\\
\indent We stress once more that the results we present here are valid in the limit $x^2\ll J/t$. In this limit the low lying charge excitations are dominated by the curvature near the band bottom of the bare dispersion $\epsilon(k)$ and their contribution to the energy denominator in (\ref{E'}) is dominated by that of the gapped spin excitations. In the opposite limit $J/t\ll x^2$ the perturbation theory presented here is still valid, yet a crossover will take place and the asymptotic expansion (\ref{Ex}) will not hold (see appendix \ref{convapp}). \\
\indent With this in mind, the first and second order energy corrections are dominated by the following terms:

\begin{equation}\label{E2_3}
\begin{split}
 E_1/L&\,\simeq \,A_1x=-\left<\gamma_j\right>_0\,Jx\\
 E_2/L&\,\simeq \,A_2x
\end{split}
\end{equation}

\noindent We will now show that the second order term is indeed suppressed by powers of $J/t$ compared to the first order term, which is of order $Jx$. To achieve a systematic expansion of $A_2$ in $J/t$ we rewrite (\ref{A2}) in the form 

\begin{equation}\label{A2_3}
  A_2=-\int_0^{2\pi}\frac{dk}{2\pi}\int dE\frac{A(k,E)}{\epsilon(k)-\epsilon(0)+E-E_{\sigma_0}}
\end{equation}

\noindent where we have introduced a spectral function 

\begin{equation}\label{A}
\begin{split}
 A(k,E)&=\frac{J^2}{N_e}\sum'_\sigma\,\left|\left<\sigma_0\left|\gamma_k\right|\sigma\right>\right|^2\delta(E_\sigma-E)\\
 &\equiv \sum_n \,K_n(E) \,e^{ikn}
\end{split}
\end{equation}

\noindent and its energy dependent Fourier coefficients $K_n(E)$. In terms of the latter we may write

\begin{equation} \label{A2_4}
  A_2=-\frac{1}{2t}\int dE\sum_nK_n(E)\int_0^{2\pi}\frac{dk}{2\pi}\frac{e^{ikn}}{1+\frac{E-E_{\sigma_0}}{2t}-\cos(k)}
\end{equation}

\noindent The $k$-integral is readily performed to give

\begin{equation} \label{A2_5}
\begin{split}
 & A_2=-\frac{1}{2t}\int dE\sum_nK_n(E)\frac{\left(1+\Delta-\sqrt{(1+\Delta)^2-1}\right)^{|n|}}{\sqrt{(1+\Delta)^2-1}}\\
&\text{where}\quad \Delta\equiv \frac{E-E_{\sigma_0}}{2t}
\end{split}
\end{equation}

\noindent The matrix elements defining $K_n(E)$ will decay rapidly when $E$ is a few times $J$ and hence we may expand (\ref{A2_5}) in powers of $\Delta$. Keeping only the leading term this yields
\begin{equation} \label{A2_final}
\begin{split}
 A_2&\simeq-\frac{1}{2t}\int dE\sum_nK_n(E)\frac{1}{\sqrt{\frac{E-E_{\sigma_0}}{t}}}\\
 &\simeq-\frac{t}{2}\Gamma_\frac{1}{2}\left(\frac{J}{t}\right)^\frac{3}{2}
\end{split}
\end{equation}

\noindent where the coefficient $\Gamma_\frac{1}{2}$ is a quantity defined only in terms of eigenstates of the doped spin chain. For later convenience we define the more general function:

\begin{equation}\label{Gamma}
\begin{split}
\Gamma_p&=J^{p-2}\int dE\,\frac{A(q=0,E)}{\left(E-E_{\sigma_0}\right)^p} \\
&=\frac{1}{N_e}\sum_{\left|\sigma\right>}'\frac{\left<\sigma_0\right|\gamma_{q=0}\left|\sigma\right>\left<\sigma\right|\gamma_{q=0}\left|\sigma_0\right>}{\left(\frac{E_\sigma-E_{\sigma_0}}{J}\right)^p} 
\end{split}
\end{equation} 

\noindent Hence, it is apparent from (\ref{E2_3}) and (\ref{A2_final}) that 

\begin{equation}\label{converge}
\frac{E_2}{E_1}\sim \left(\frac{J}{t}\right)^\frac{1}{2}
\end{equation}

\noindent indicating the convergence of our perturbative approach for small $J/t$. In appendix \ref{convapp}, we will further comment on convergence and expansion parameters of this series in various limits. Note that the non-analytic nature of the expansion originates from the gaplessness of the charge degrees of freedom and the existence of a regime where the spin gap dominates the energy denominator in (\ref{A2_3}). Similarly, in second order perturbation theory the compressibility parameter $v_N$ reads to leading order in $x$:

\begin{equation}\label{vn_2}
v_N\simeq \left(4\pi t+\frac{12}{\pi}C_2\right) x
\end{equation}

\begin{figure}[hbt]
\samepage
\includegraphics[width=3.3in]{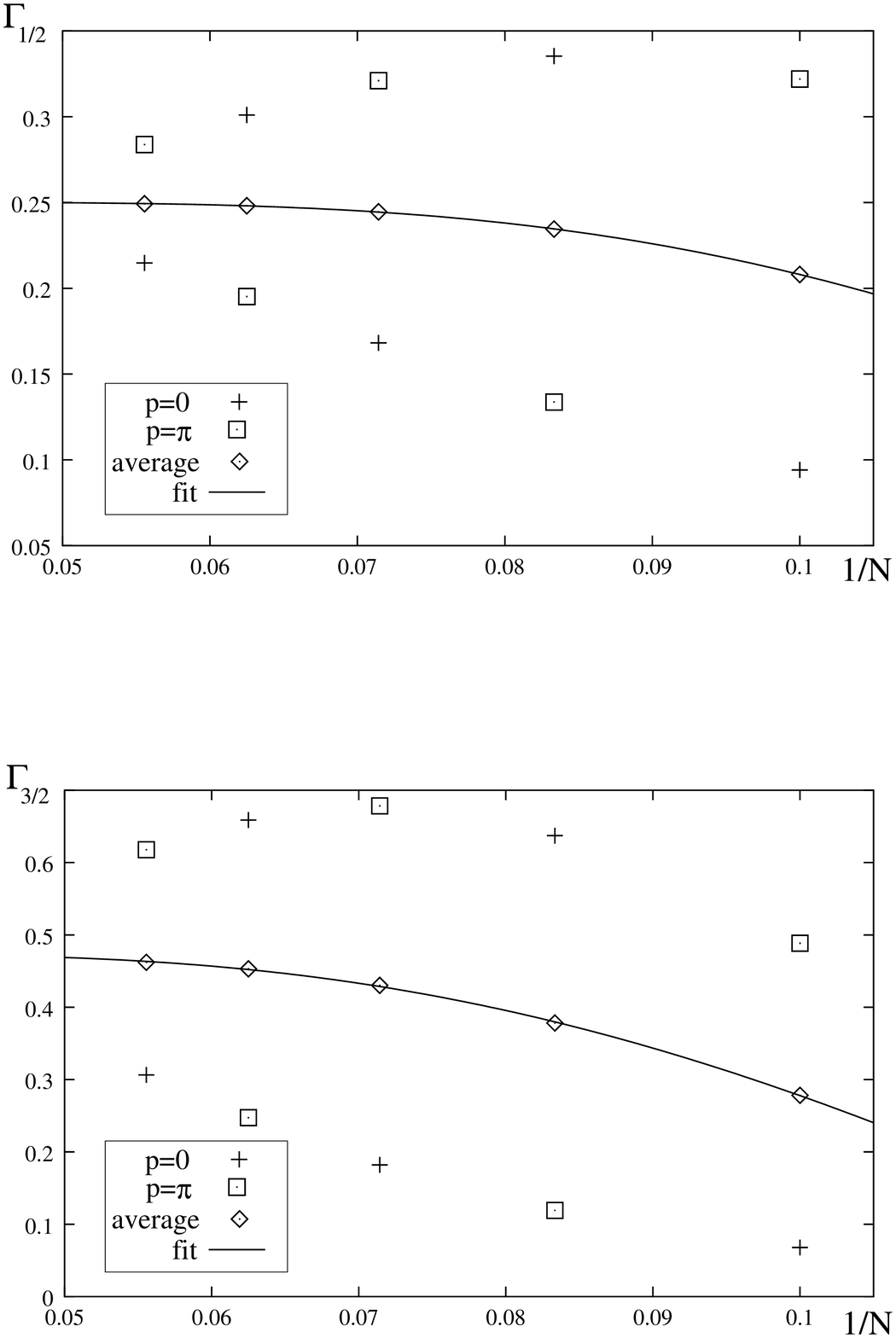}
\caption{\label{gammas} Numerical determination of $\Gamma_\frac{1}{2}$ and $\Gamma_\frac{3}{2}$ by exact diagonalization of $H_s$ for $\alpha=\frac{1}{2}$. System sizes of up to $N=18$ have been diagonalized. Results are plotted for the two degenerate ground states with momenta $p=0$ (crosses) and $p=\pi$ (squares). The extrapolated values have been determined by fitting their averages (diamonds) to the function $f(N)=a+b\,\exp(-cN)$}. 
\vspace{-.5cm}
\end{figure}

\noindent By means of (\ref{C2_2}) the evaluation of $C_2$ goes analogous to that of $A_2$ and we get: 
\begin{equation}\label{C2_final}
\begin{split}
C_2&\simeq-\frac{\pi^2}{12}\Gamma_\frac{3}{2}\sqrt{Jt}\\
v_N&\simeq\pi t\left(4-\Gamma_\frac{3}{2}\sqrt{\frac{J}{t}}\right)x
\end{split}
\end{equation}
\noindent Numerically, we found $\Gamma_\frac{1}{2}=.2502(2)$ and $\Gamma_\frac{3}{2}=.474(3)$ at $\alpha=.5$ (Fig. \ref{gammas}). Hence, although the compressibility $\kappa\sim v_N^{-1}$ increases with $J$, no unstable value of $J$ can be inferred that lies within the validity of our perturbation theory.\\

\section{\label{polaron} Single spin-polaron picture}

\noindent We will now develop a variational picture of the polaronic effects of a single hole on its spin environment at small $J/t$ in the special case $\alpha=.5$. The perturbation theory presented in the preceding section for a finite carrier concentration may be applied to the problem of doping a single hole into the infinite spin chain as well, such that we will be able to compare variational and perturbative results. In second order perturbation theory, the energy of a single hole at momentum $k$ reads:

\begin{multline}\label{Ep}
   E_p(k)= -2t\cos(k)-\left<\gamma_j\right>_0 J
          - \int_{0}^{2\pi}\frac{dk_2}{2\pi}\;f(k,k_2)+\dotsc 
\end{multline}

\noindent where we have not included the contribution $E_{\sigma_0}$ from the spin background. At $k=0$ we immediately see by comparison with (\ref{zero}),(\ref{A1}) and (\ref{A2}) that

\begin{equation}\label{Ep_2}
   E_p\equiv E_p(k=0) = A_0 +A_1 +A_2+\dotsc \equiv A 
\end{equation}

\noindent holds for the single polaron energy in second order perturbation theory. Likewise, for the renormalized mass of the spin polaron we have, comparing to eqs. (\ref{zero}) and (\ref{C2_2})

\begin{equation}\label{m}
\begin{split}
  m^{-1}\equiv \left.\frac{\partial^2}{\partial k^2}E_p\right|_{k=0} = \frac{6}{\pi^2}\left(C_0+C_2+\dotsc\right)\equiv \frac{6}{\pi^2}C\end{split}
\end{equation}

\noindent at this order. We may therefore rewrite the ground-state energy of the system at finite doping (\ref{Ex}) as 

\begin{equation}
\label{E_non_int}
\begin{split}
E&=E_{\sigma_0}+L\left(E_px +\frac{\pi^2}{6m}x^3+\dotsc\right)\\
 &= E_{\sigma_0}+N_e\int_{-k_f}^{k_f}\frac{dk}{2\pi}\,E_p(k)\;+O(k_f^4)
\end{split}
\end{equation}

\noindent Hence up to third order in $x$ the ground-state energy of the system is apparently given by the energy of non-interacting spinless particles with a dispersion $E_p(k)$, where interaction effects enter only beyond this order. This further confirms the picture established in the preceding sections.\\
\indent We now focus on the Majumdar-Gosh point $\alpha=.5$, where the ground state of the spin Hamiltonian $H_s$ is known exactly \cite{MG}. It consists of a direct product of uncorrelated singlet pairs:

\begin{equation}\label{MGGS}
 \left|MG\right>=\prod_i\frac{1}{\sqrt{2}}\Bigl(\left|\uparrow\downarrow\right>- \left|\downarrow\uparrow\right> \Bigr)_{2i,2i+1}
\end{equation}

\begin{figure}[hbt]
\samepage
\vspace{.2cm}
\includegraphics[width=3.3in]{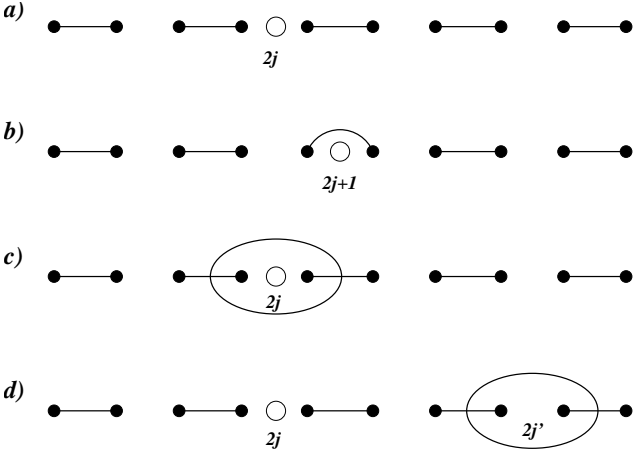}
\caption{\label{states2} a)+b) Single hole basis states forming the ground state of the non-interacting Hamiltonian $H_0$ (eq.(\ref{GS0})). Lines denote the singlet pairs in (\ref{MGGS}). c) A pair of triplets excited by the presence of the hole. The oval denotes a singlet formed by two triplet states on the links adjacent to the hole, as displayed in (\ref{triplex}). d) States used to form the variational wave function (\ref{t}).}   
\vspace{-.5cm}
\end{figure}

\noindent 
Note that we use $|MG>$ to denote one of the two doubly degenerate symmetry broken ground states, whereas $|\sigma_0>$ has been used to denote their symmetric superposition. At $\alpha=.5$, our results for the single polaron energy and mass eqs. (\ref{Ep_2}) and (\ref{m}) take the concrete form
\begin{equation}\label{E&M}
\begin{split}
E_p &=-t\left( 2-\frac{9}{16}\frac{J}{t}+.125 \left(\frac{J}{t}\right)^\frac{3}{2}+\dotsc\right)\\
m^{-1}&=t\left(2-.237\sqrt{\frac{J}{t}}+\dotsc\right)
\end{split}
\end{equation}
\noindent 
We may write the ground state of the unperturbed Hamiltonian $H_0$ as a superposition of states depicted in Fig. \ref{states2}a)+b):
\begin{equation}\label{GS0}
\left|\Omega_0\right> = \frac{1}{\sqrt{N_e}}\sum_j\left|j\right>_c\left|MG\right>_s
\end{equation}
\noindent where $\left|j\right>_c$ denotes a state with a hole at the interstitial site $j$ and ``$c$'' and ``$s$'' refer to the spin sector and the charge sector of the state. When the interaction $H_{sc}$ is taken into account, a hole in the state $\left|2j\right>_c$ will excite a spin configuration where the two dimers adjacent to the hole are in triplet states, and the two triplets form
 a singlet (Fig. \ref{states2}c)). More precisely, the oval in Fig. \ref{states2}c) denotes the following spin state:
\begin{equation}\label{triplex}
\begin{split}
&\mbox{\includegraphics[width=1.1in]{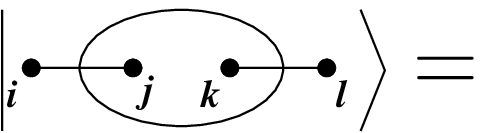}}\\
&\quad\frac{1}{\sqrt{3}}\biggl(\left|\uparrow\uparrow\right>_{ij}\otimes \left|\downarrow\downarrow\right>_{kl}
                         +\left|\downarrow\downarrow\right>_{ij}\otimes\left|\uparrow\uparrow\right>_{kl}\biggr.\\
       &\quad\quad \quad  -\frac{1}{2}\Bigl( \left|\uparrow\downarrow\right>+\left|\downarrow\uparrow\right>\Bigr)_{ij}\otimes
                             \biggl.  \Bigl( \left|\uparrow\downarrow\right>+\left|\downarrow\uparrow\right>\Bigr)_{kl}\biggr)
\end{split}
\end{equation}
\noindent We may now denote such a triplet pair excitation which is centered around the hole site $2j$ by $\left|2j\right>_s$. Similarly, the hole state  $\left|2j+1\right>_c$ will excite the spin states  $\left|2j\right>_s$ and  $\left|2j+2\right>_s$. Clearly, a variational wave function will need admixtures of states such as shown in  Fig. \ref{states2}c). However, in the vicinity of the hole the kinetic energy $H_c$ is the dominant part of the Hamiltonian, and it will allow the hole to move away from the excited triplet states as shown in Fig. \ref{states2}d). To optimize the kinetic energy, it is hence necessary to include the more general states of Fig. \ref{states2}d) into the wave function.
We therefore write down the following trial wave function for a state with one hole at momentum $k$:

\begin{multline}\label{t}
  \left|t_k\right>=\frac{1}{\sqrt{N_e}}\sum_je^{ikj}\left|j\right>_c\otimes\\
\left(\left|MG\right>_s+a\sum_{2j'}e^{-\left|j-2j'\right|\beta+i\left(j-2j'\right)\delta}\left|2j'\right>_s\right)
\end{multline} 

\noindent Hence $\left|t_k\right>$ lives in the subspace of all states that can be reached by acting once with the perturbation $H_{sc}$ on the zeroth order wave function (\ref{GS0}) and then acting an arbitrary number of times with the hopping operator $H_c$. We also note that indeed certain exact excited states of a Majumdar-Gosh spin chain are given in terms of the tightly bound triplet excitations shown in eq. (\ref{triplex}) and Fig. \ref{states2} (Ref. \onlinecite{CASPERS}). The following matrix elements are needed to evaluate the energy of the state (\ref{t}):
\begin{align}
_{\substack{\\s\hspace{-.05cm}}}\left<MG\left|-J\,\gamma_j\right|2j'\right>_s&=\frac{\sqrt{3}}{8}J\left(\delta_{j,2j'}+\delta_{j-1,2j'}+\delta_{j+1,2j'}\right)\label{mat1}\\
_{\substack{\\s\hspace{-.05cm}}}\left<2j'\left|-J\,\gamma_j\right|2j'\right>_s&=\frac{9}{16}J-(-1)^j\frac{3}{16}J\nonumber\\
&+\frac{1}{4}J\left(3\delta_{j,2j'}-\delta_{j-1,2j'}-\delta_{j+1,2j'}\right)\label{mat2}\\
_{\substack{\\s\hspace{-.05cm}}}\left<2j'\left|H_{s}\right|2j'\right>_s&= \;_{\substack{\\s\hspace{-.05cm}}}\left<MG\left|H_{s}\right|MG\right>_s+J\nonumber\\
&=\;-\frac{3}{4}JN_e\;+\;J\label{mat3}
\end{align} 
\noindent In addition, both $\gamma_j$ and $H_s$ do not have off-diagonal matrix elements among the states $\left|2j\right>_s$. This leads to the following expectation values:

\begin{subequations}\label{expect}
\begin{align}
  \left<t_k|H_c|t_k\right>&= -2t\cos(k)-4t|a|^2 \xi_3 \cos(k+\delta)\label{Hcex}\\
  \left<t_k|H_s|t_k\right>&=-\frac{3}{4}JN_e \left<t_k|t_k\right>+J|a|^2(\xi_1+\xi_2)\label{Hsex}\\
  \left<t_k|H_{sc}|t_k\right>&=\frac{9}{16}J\left<t_k|t_k\right>\nonumber\label{Hscex}\\
 &+\frac{\sqrt{3}}{16}J\left(1+2e^{-\beta}\cos(\delta)\right)(a+a^*)+O(J|a|^2)\\
\left<t_k|t_k\right>&=1+|a|^2(\xi_1+\xi_2)
\end{align}
\end{subequations}

\noindent where the constants $\xi_1$ and $\xi_2$ are proportional to the weight of spin excited states with the hole on even positions and odd positions, respectively, and $\xi_3$ arises from hopping between even and odd sites in the presence of a spin excitation:

\begin{equation}\label{xis}
\begin{split}
\xi_1&=\frac{1}{2}\sum_{j'}e^{-2\beta|2j'|}=\frac{1}{4\beta}+\frac{1}{3}\beta +\dotsc\\
\xi_2&=\frac{1}{2}\sum_{j'}e^{-2\beta|2j'-1|}=\frac{1}{4\beta}-\frac{1}{6}\beta +\dotsc\\
\xi_3&=\frac{1}{4}\sum_{j'}e^{-\beta|2j'|}\left(e^{-\beta|2j'-1|}+e^{-\beta|2j'+1|} \right)=\frac{1}{4\beta}-\frac{1}{24}\beta +\dotsc
\end{split}
\end{equation}

\noindent Terms of order $J|a|^2$ were only kept in (\ref{expect}) when they are multiplied by $\xi_i\sim 1/\beta$. It is apparent from (\ref{Hcex}) that $\delta=-k$ has to be chosen, and from (\ref{Hscex}) that $a$ is real and negative. Keeping only leading terms, this leads to the variational energy function

\begin{equation}\label{Evar}
\begin{split}
E_p^{var}(k;a,\beta)&\equiv\left<t_k|H_c+H_s+H_{sc}|t_k\right>/\left<t_k|t_k\right>\\
&=-2t\cos(k)+\frac{9}{16}J
+\frac{1}{2}ta^2\beta-t\frac{a^2}{\beta}(1-\cos(k))\\
&\quad+J\frac{a^2}{2\beta}
   +\frac{\sqrt{3}}{8}Ja\left(1+2\cos(k)\right)
\end{split}
\end{equation}

\noindent where again the bulk contribution of the spin chain was not included. We first minimize this function for $k=0$ and find for the variational parameters at the stationary point

\begin{subequations}\label{var}
\begin{align}
  \beta_0&=\sqrt{\frac{J}{t}}\label{a0}\\
   a_0&=-\frac{3\sqrt{3}}{16}\sqrt{\frac{J}{t}}
\end{align}  
\end{subequations}

\noindent By (\ref{a0}), the size of the spin polaron cloud is proportional to $\left(J/t\right)^{-\frac{1}{2}}$ in agreement with (\ref{Ep}) where the dominant contributions to the integral come from the region where $k_2^2$ is of the order of the spin gap. The variational energy of the spin polaron at $k=0$ is thus 

\begin{equation}\label{Evar_2}
E_p^{var}\equiv E_p^{var}(k=0;a_0,\beta_0)=-t\left( 2-\frac{9}{16}\frac{J}{t}+\frac{27}{256}\left(\frac{J}{t}\right)^\frac{3}{2}\right)
\end{equation}

\noindent This is indeed of the same form as (\ref{E&M}) where the first two terms are reproduced exactly, as they are mean-field like in character. Moreover, the coefficient of the last term is about $.105$ and hence matches the one obtained by perturbative and numerical methods in (\ref{E&M}) within roughly $15\%$. \\
The appearance of a mass term proportional to $\sqrt{J/t}$ as in (\ref{E&M}) may also be understood from this variational approach. It is seen in (\ref{Evar}) that a term of order $a^2/\beta\sim\sqrt{J}$ is no longer precisely canceled at finite $k$. The reason for this is that at finite $k$ time reversal symmetry is absent and a non-zero value of the parameter $\delta$ introduced in (\ref{t}) is generally allowed. We have tuned $\delta$ such that the polaronic corrections in the kinetic energy (\ref{Hcex}) do not have the same $k$-dependence as the leading term. This is giving rise to a $a^2/\beta$ term at finite $k$. It leads to the variational mass

\begin{equation}
\begin{split}
  \left(m^{var}\right)^{-1}&=\left.\frac{\partial^2}{\partial k^2} E_p^{var}(k;a_0,\beta_0)\right|_{k=0}\\
&=t\left(2-\frac{27}{256}\sqrt{\frac{J}{t}}+\dotsc\right)
\end{split}
\end{equation}

\noindent Here, the dependence of $a$ and $\beta$ on $k^2$ need not be taken into account because of stationarity. The coefficient of the second term happens to be the same as the one showing up in (\ref{Evar_2}) which is now off by about a factor of $2$ when compared to the mass shown in (\ref{E&M}). This may be attributed to the variational character of the state (\ref{t}), since the mass comes from a subdominant term proportional to $k^2$. However, the correct dependence on $J$ as well as the right order of magnitude are again obtained. We therefore conclude that the wave function (\ref{t}) provides a quite accurate picture of the large polaronic cloud in the limit of small $J/t$, especially at $k=0$.\\

In view of our original motivation to examine the stability of the liquid phase of the $t$-$J$-$J'$ model as $x\rightarrow 0$ at small $J/t$, it is interesting to think about the possibility of the formation of bound hole states. It is generally expected that either at the critical value for the onset of phase separation, $(J/t)_c$, or at an even smaller critical value ${(J/t)_{c_1}}<(J/t)_c$ bound states of two holes will exist \cite{HALCOM} (see also Ref. \onlinecite{OGSOAS}). The existence of such bound states can be discussed on a qualitative level based on the variational spin polaron picture proposed in this section. To form a bound state, the single polaron wave functions must significantly overlap, hence the size of a bound state will be of order $r\sim (J/t)^{-\frac12}$. The potential energy gain will be of order $(J/t)^{\frac{3}{2}}$ since the mean field term of order $J$ in (\ref{Evar_2}) will not be affected by pair formation. However, the kinetic energy cost of such a state is of order $1/r^2\sim J/t$ and is dominant. We conclude that bound states of holes will require finite $J/t$ of order $1$ or greater, in agreement with the picture of free single hole-like charge degrees of freedom established in the preceding sections.

\section{\label{conclusion}Conclusion}

\noindent We have examined the $t$-$J$-$J'$-model in one dimension, in the regime of small $x$ and $J/t$ by perturbative and variational approaches. This parameter regime is most challenging to numerical methods, and earlier numerical studies did not allow a firm conclusion whether a phase separation instability and a phase of dominant singlet superconducting fluctuations extend down to values of $J/t<1$ in the case $\alpha=J'/J\approx .5$, where a spin gap is present at small doping.\\ \indent Using an approach where couplings between spin and charge degrees of freedom are treated as a perturbation, we have presented a detailed analysis of the model in second order perturbation theory, showing that no instability is present at small $J/t$. Instead, using Luttinger liquid arguments and by studying the dispersion of a single hole immersed into the correlated spin system, we have demonstrated that the hole degrees of freedom precisely behave as free spinless solitons in the dilute limit, despite their microscopic coupling to the non-trivial spin background. This behavior conforms to Luttinger liquid physics, where spin and charge are separate degrees of freedom, and couplings between them are regarded as irrelevant in a renormalization group sense. While this point of view is generally accepted for one-dimensional systems, in microscopic one-dimensional lattice models it usually may be firmly demonstrated only at special integrable points \cite{HAL1,SCHULZ}. The method we established in section \ref{pert} provides a perturbative framework for such a demonstration in a non-integrable model over a range of parameters. Moreover, it allows the calculation of non-trivial quantities such as the leading corrections to the single hole energy and mass renormalization, which depend on non-analytic powers of $J/t$. The numerical calculation of the coefficients in this expansion still requires an exact diagonalization of a pure spin problem. We used these results for a comparison to a variational approach. Proposing a variational wave function where the hole is surrounded by a polaronic cloud of tightly bound pairs of triplet excitations we were able to confirm the perturbative results for the dependence of the single polaron energy and mass on $J/t$, as well as the order of magnitude of the coefficients. In particular, the second order perturbative energy corrections are in close quantitative agreement with the variational result. Based on these findings, we argue that for the parameter $\alpha=.5$ the onset of phase separation at small doping as well as the formation of bound states require $J/t$ to be at least of order $1$.\\
\indent To conclude, the theory of the $t$-$J$-$J'$-model presented here is limited to small values of $J/t$ and cannot access a region of dominant superconducting correlations, which at $\alpha=.5$ might exists at moderate values of $J/t$ and small doping from numerics. In a real system, additional interchain effects must be taken into account, that can either favor dimer locking, or a superconducting dimer liquid when dimer locking is frustrated (cf. Ref. \onlinecite{TIOCL}). 
This calls for further investigations.

\begin{acknowledgments}
We thank F.D.M. Haldane for insightful discussions. This work was supported by the MRSEC program of the NSF under award number DMR 0201069.  
\end{acknowledgments}

\appendix

\section{\label{convapp}Convergence and crossover behavior of the perturbative expansion}

\noindent Here we briefly illustrate the behavior of our expansion at $k$'th order perturbation theory, where one will encounter terms analogous to (\ref{E'}):

\begin{equation}\label{Ek}
\begin{split}
   E_k&\sim \int_{-k_f}^{k_f}dh_1\dotsc\int_{-k_f}^{k_f}dh_m\int_{k_f}^{2\pi-k_f}dp_1\dotsc\int_{k_f}^{2\pi-k_f}dp_n \\
&\sum_{\sigma_1\dotsc \sigma_{k-1}}
\underset{\text{\small $k-1$\;\;\rm factors}}{\underbrace{\frac{J^kM\left(h_1\dotsc h_m;p_1\dotsc p_n;\sigma_1\dotsc\sigma_{k-1}\right)}{\bigl(\epsilon(p_1)-\epsilon(h_1)+E_{\sigma_1}-E_{\sigma_0}\bigr)\bigl(\quad\bigr)\dotsc\bigl(\quad\bigr)}}}+\dotsc 
\end{split}
\end{equation}  

\noindent The phase space consists of $m$-hole momenta and $n$-particle momenta. It is enough to consider the case $m+n=k$. There will be terms with fewer integrals also, but they are multiplied by additional powers of $x$ such as (\ref{E0}). \\
\indent We focus on the regime $k_f^2\ll J/t$ first.
Since $\epsilon(k)\approx {\rm const}+tk^2$, the integrand does not significantly depend on the hole momenta such that each of the hole integrals will give rise to a factor of $x$. The integral over particle momenta $p_i$ will be dominated by the region where all momenta are within a range of $\sqrt{\Delta/t}$ of the Fermi points, where $\Delta\sim J$ is the spin gap. In this regime, all of the $k-1$ factors in the denominator are dominated by the spin gap and are of order $\Delta$. Hence we obtain the following estimate for the term displayed in (\ref{Ek})

\begin{equation}\label{est}
 \left(\ref{Ek}\right)\sim x^m\left(\Delta/t\right)^{n/2} \frac{J^k}{\Delta^{k-1}}\sim \;x^m\left(\sqrt{J/t}\right)^nJ
\end{equation}

\noindent The leading contribution to $E_k$ in the limit $x^2\ll J/t$ will thus be a term of order

\begin{equation}\label{Ek2}
 E_k\sim \;x\left(\sqrt{J/t}\right)^{k-1}J
\end{equation}

\noindent Eq. (\ref{Ek2}) shows that subsequent orders in perturbation theory are always suppressed by powers of $\sqrt{J/t}$, as we verified explicitly up to second order (cf. (\ref{converge})). Note that relations (\ref{est}) and (\ref{Ek2}) are valid asymptotically in a given limit, they do not imply the existence of a systematic expansion in powers of $x$ and $\sqrt{J/t}$. Rather, the $E_k$'s are quite complicated functions of $x$ and $J$. Relation (\ref{Ek2}) will hold until $x^2/J\sim 1/t$, and upon further increase of this ratio a crossover will take place. We may however write down an asymptotic expansion in $x$:
\begin{equation}
\label{Ek3}
\begin{split}
E_k&= L\left(A_k(J)x +B_k(J)x^2 +C_k(J)x^3+D_k(J)x^4\dotsc\right)\\
E\,&=\text{const.}+E_1+E_2+\dotsc
\end{split}
\end{equation}

\noindent as we did in second order perturbation theory. Recall from (\ref{A2_final}), (\ref{C2_final}) that $A_2\sim (J/t)^{3/2}$, $C_2\sim (J/t)^{1/2}$ while $B_2=0$. Formally, however, $B_2$ is of order $J$. This implies that in general $\sqrt{tx^2/J}$ is the expansion parameter of the series (\ref{Ek3}).\\
\indent In the opposite limit $J\ll tx^2$ it is easily seen from (\ref{Ek}) that now $E_k\sim J^k$ holds. In this limit it is not necessary, though still permissible, to include the spin chain part $H_s$ in the unperturbed Hamiltonian $H_0$. Instead one may apply degenerate perturbation theory in the spin couplings, which gives rise to an asymptotic expansion in $J$:

\begin{equation}
\label{Ek4}
E= \text{const}+L\left(a(x)J +b(x)J^2 +c(x)J^3+d(x)J^4\dotsc\right)
\end{equation}

\noindent This method has been applied in Ref. \onlinecite{OGLURI} to calculate the intersection of the spin gap phase boundary with the $x$-axis. Note that in (\ref{Ek4}) $x$ need not be small, whereas in (\ref{Ek3}) both $J/t$ and $x^2t/J$ have to be small. However, due to the limitation $J\ll tx^2$, eq. (\ref{Ek4}) cannot be used to address the nature of the phase diagram in the dilute hole limit.

\section{\label{Fcontin} The continuity of the function $\pmb F(q)$}

\noindent We will now give an argument for the continuity of the function $F(q)$ which leads to the crucial cancellation in (\ref{B2_2}). This question is more subtle than it may seem, and the following argument would require more scrutiny in the gapless case $\alpha<\alpha_c$. We restrict ourselves to the spin gapped case, as we have done throughout the paper. Recall the definition of $F(q)$ from (\ref{F}):\\

\begin{equation}\label{Fapp}
\begin{split}
  F(q)&=\frac{J^2}{2N_e}\left(\sum_{\left|\sigma\right>}'\frac{\left<\sigma_0\right|\gamma_{-q}\left|\sigma\right>\left<\sigma\right|\gamma_{q}\left|\sigma_0\right>}{E_\sigma-E_{\sigma_0}}+(q\rightarrow -q)\right)
\end{split}
\end{equation}

\indent Physically, the continuity of $F(q)$ can be seen by interpreting $F(q)$ as the second order energy response of a pure spin chain due to a periodic perturbation. More precisely, we consider the following auxiliary spin chain problem:
\begin{equation}\label{Haux}
\begin{split}
  H_q(\lambda)&=H_s+\lambda\,J\sum_j\cos(qj)\gamma_j\\
              &=H_s+\frac{\lambda}{2}\,J\left(\gamma_q+\gamma_{-q}\right)
\end{split}
\end{equation}

\noindent where $H_s$ is as defined in (\ref{Hs}). Let ${\cal E}_q(\lambda)$ denote the ground-state energy per site of this problem. Then it is easily seen from second order perturbation theory and the definition (\ref{Fapp}) that at $q=0$
\begin{equation}\label{q0}
  F(0)=-\frac{1}{2}\,{\cal E}''_{q=0}
\end{equation}
\noindent holds, where the prime denotes a derivative with respect to $\lambda$ taken at $\lambda=0$. On the other hand, at $q\neq 0$ the same argument gives
\begin{equation}\label{qneq0}
  F(q\neq 0)=-\,{\cal E}''_{q}
\end{equation}

\noindent Note the factor of $2$ difference between (\ref{q0}) and (\ref{qneq0}). Despite this apparent difference between the cases $q=0$ and $q\neq 0$, it is ${\cal E}''_q$ which is discontinuous at $q=0$, not $F(q)$, as the following argument shows: In the vicinity of a site $j$ the ground state of $H_q(\lambda)$ will have great overlap with the ground state of $H_{q=0}(\lambda(j))$ as $q\rightarrow 0$, where $\lambda(j)\equiv \lambda \cos(qj)$.  In other words, as $q\rightarrow 0$ it should be justified to replace the oscillating perturbation in $H_q(\lambda)$ by a flat perturbation in a sufficiently large local region around each site $j$. The size of this region can still be chosen to be $\ll 1/q$. 
One can thus  argue that up to powers of $q$ the ground-state energy will be given by a sum over local contributions ${\cal E}_{q=0}(\lambda(j))$:

\begin{equation}\label{limq0}
\begin{split}
  {\cal E}&_{q\rightarrow 0}(\lambda)=\frac{1}{N_e}\sum_j {\cal E}_{q=0}(\lambda(j)) 
   \approx \frac{1}{N_e} \int_0^{N_e}dx \;{\cal E}_{q=0}(\lambda\cos(qx)) \\
&= \frac{1}{N_e} \int_0^{N_e}dx \left({\cal E}_{q=0} +\lambda\cos(qx){\cal E}'_{q=0} +\frac{1}{2}\lambda^2\cos^2(qx){\cal E}''_{q=0} +\dotsc\right)\\
&\approx \;{\cal E}_{q=0} + \frac{1}{4}\lambda^2{\cal E}''_{q=0}
\end{split}
\end{equation}
From (\ref{qneq0}), (\ref{q0}) it then follows that 

\begin{equation}\label{Fcont}
  F(q\rightarrow 0) = -\frac{1}{2} {\cal E}''_{q=0}= F(0)
\end{equation}

\noindent Note that the local point of view taken here is better justified in the gapped case, where any local perturbation decays exponentially in space.


\end{document}